\documentclass[twocolumn,aps,preprintnumbers,prl,showpacs,amsfonts,amsmath,amssymb,superscriptaddress,floatfix]{revtex4}

\usepackage[dvips]{graphicx}
\usepackage[tight]{subfigure}
\usepackage{times}

\allowdisplaybreaks[3]

\newcommand{\hatd}[1]{\hat{#1}^\dagger}

\newcommand{\ket}[1]{ |{#1} \rangle}
\newcommand{\expectation}[1]{\langle #1 \rangle}

\newcommand{\ox}{\hat{x}}
\newcommand{\op}{\hat{p}}

\newcommand{\affA}{%
    Department of Applied Physics, School of Engineering,
        The University of Tokyo,\\
    7-3-1 Hongo, Bunkyo-ku, Tokyo 113-8656, Japan}

\newcommand{\affB}{%
Department of Optics, Palack\'{y} University, 17. listopadu 50, 772 07 Olomouc, Czech Republic
}

\newcommand{\affC}{%
    Optical Quantum Information Theory Group,
Max Planck Institute for the Science of Light,\\
Institute of Theoretical Physics I, Universit\"{a}t Erlangen-N\"{u}rnberg,
Staudtstr.7/B2, 91058 Erlangen, Germany}

\begin{document}

\title{Demonstration of Cluster State Shaping and Quantum Erasure for Continuous Variables}

\date{\today}

\author{Yoshichika Miwa}
\affiliation{\affA}
\author{Ryuji Ukai}
\affiliation{\affA}
\author{Jun-ichi Yoshikawa}
\affiliation{\affA}
\author{Radim Filip}
\affiliation{\affB}
\author{Peter van Loock}
\affiliation{\affC}
\author{Akira Furusawa}
\affiliation{\affA}


\begin{abstract}
We demonstrate experimentally how to remove an arbitrary node from a continuous-variable cluster state and how to shorten any quantum wires of such a state.
These two basic operations, performed in an unconditional fashion,
are a manifestation of quantum erasure and can be employed to obtain various graph states from an initial cluster state.
Starting with a sufficiently large cluster, the resulting graph states can then be used for universal quantum information processing. 
In the experiment, all variations of this cluster-shaping are demonstrated on a four-mode linear cluster state through homodyne measurements and feedforward.
\end{abstract}


\maketitle


A one-way quantum computer uses a cluster state as a resource~\cite{one-wayRaussendorf,one-wayMenicucci}.
For qubits, cluster states have graph structure where the nodes are qubits while the bonds represent controlled $\pi$-phase-shift interactions~\cite{one-wayRaussendorf}.
Instead of the computational basis for a qubit ($\ket{0}$, $\ket{1}$), coordinate eigenstates $\ket{x}$ for any real number $x$ correspond to the computational basis in the continuous-variable (CV) case.
A very powerful scheme for generating a large-scale but fixed CV cluster state in just one time step
was recently proposed in Ref.~\cite{ofc.prl} using optical frequency combs. Starting with such large-scale cluster states enables one to use a {\em top-to-bottom} approach, in which
the initial state can be shaped and converted into a modified, smaller cluster state suitable for a 
given quantum computation task. In order to adapt the fixed, initial cluster state to any desired operation, 
we need to be able to flexibly shape the cluster state.

For CV cluster states, the nodes are quantized optical modes coupled through QND interactions~\cite{one-wayMenicucci}.
Once these cluster nodes are linked, they can be easily decoupled by measurements in the computational basis and feedforward, 
as we have reported for the CV case~\cite{Filip03.pra, erasing_paper}. 
This decoupling is a manifestation of a complete CV quantum erasure which works independently of the input states for this QND-type interaction.
Only two basic shaping operations are required to shrink a fixed large-scale cluster state and transform it into an appropriate form 
for a desired quantum operation~\cite{one-wayGu}.
One is the removal of unwanted nodes by measuring the modes to be removed and performing feedforward on the neighboring modes,
at the same time breaking all bonds between that node and the rest
(here feedforward corresponds to phase-space displacements depending on the homodyne measurement results).
The other operation is wire-shortening~\cite{one-wayGu}, which also removes modes from a cluster state,
but it leaves the neighboring modes connected in the resulting graph state.

In this paper, we demonstrate {\em unconditional} cluster-state shaping for CV cluster states, specifically, for a CV four-mode linear cluster state.
Similar to the qubit case, feedforward corrections are essential to accomplish the cluster shaping.
However, so far cluster-shaping has not been demonstrated in a deterministic and unconditional fashion,
since the single-photon-based qubit proposals~\cite{Varnava06} and implementations~\cite{Walther05}
are typically heralded, relying upon post-selection.
In contrast, the CV approach for creating and shaping cluster states does not require any quantum memories for storage.
The price for this unconditionalness, however, is that the finitely-squeezed CV cluster states are intrinsically
imperfect~\cite{one-wayMenicucci,clusterPeter,Eisert10}.

\begin{figure}[tb]
\subfigure[Removing a mode from a cluster state.]{
\includegraphics[clip, scale=0.3]{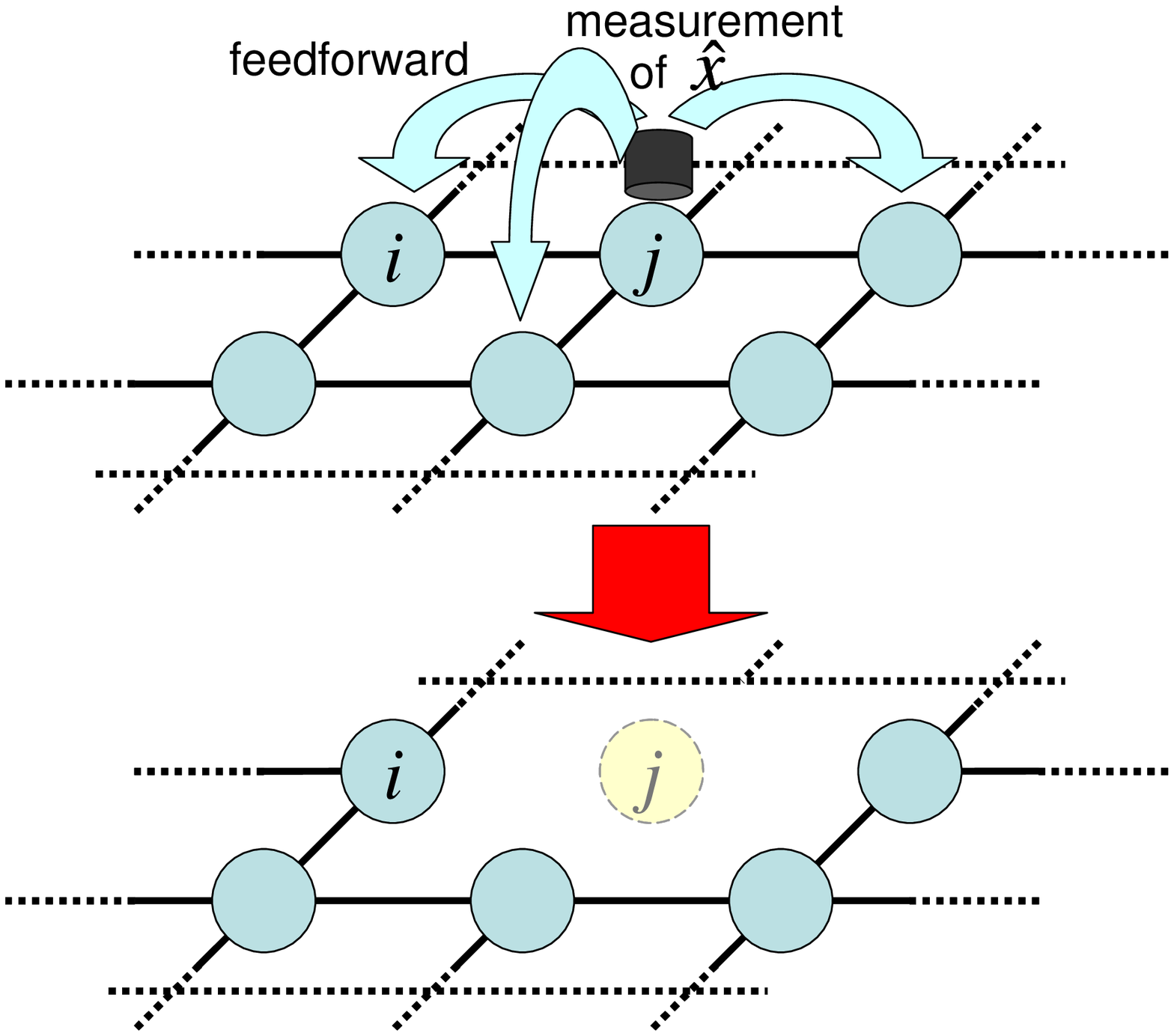}
\label{fig:removing}
}
\subfigure[Wire-shortening.]{
\includegraphics[clip, scale=0.3]{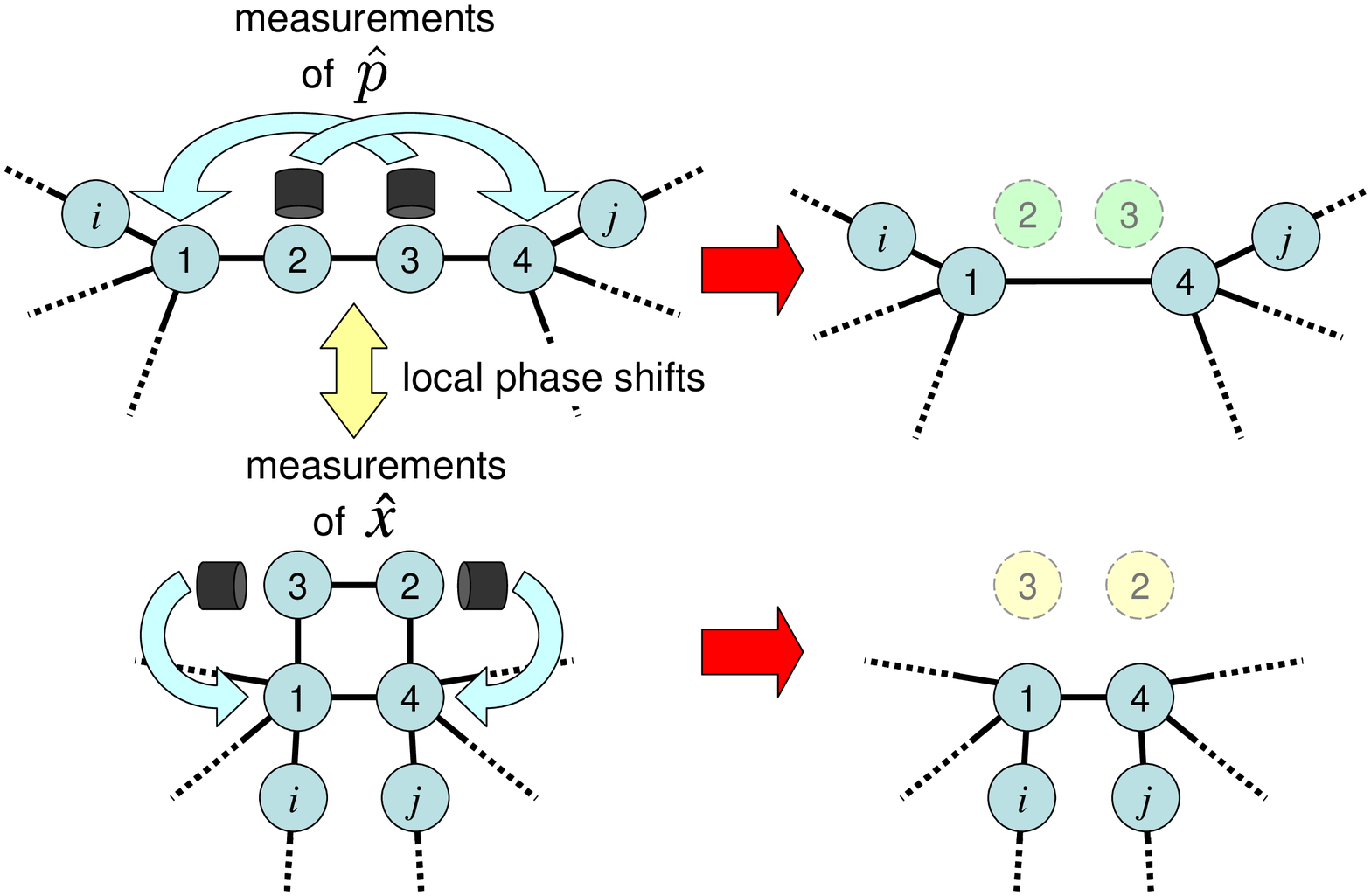}
\label{fig:wire_shortening_theory}
}
\subfigure[Realizing scheme of shaping a cluster state.]{
\includegraphics[clip, scale=0.3]{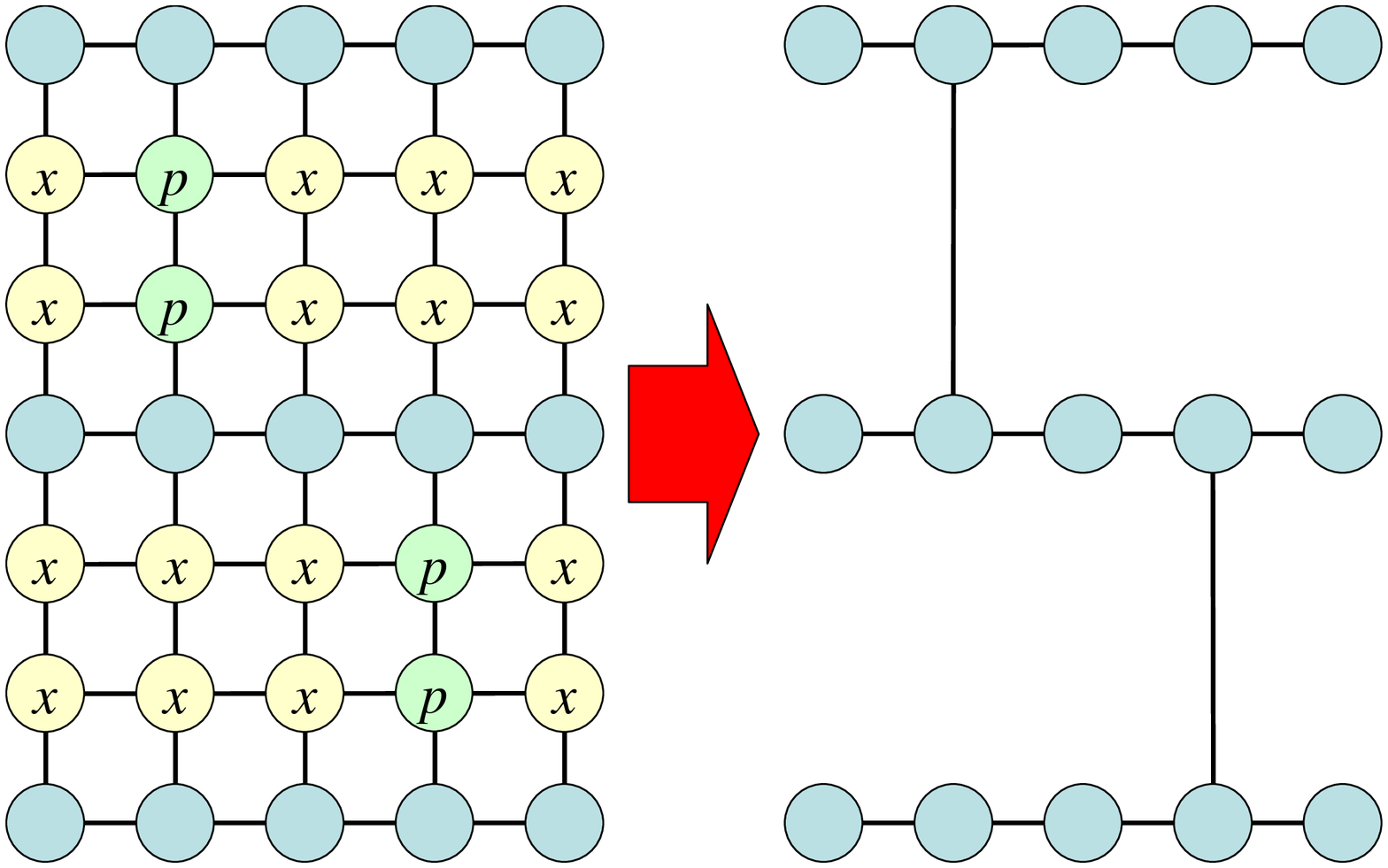}
\label{fig:cluster}
}
\caption{Cluster-state shaping using measurement and feedforward. Measuring $x$ simply remove the modes while measuring $p$ remove the modes with their quantum correlation reconstructed in their neighbors.}
\end{figure}

\textit{Theory.}--- Each quantum mode can be represented by a pair of observables $\ox$ and $\op$ in Heisenberg picture, where these operators are the real and imaginary parts of each mode's annihilation operator, $\hat{a} = \ox +i\op$.
The commutation relation of these operators is $[\ox_i, \op_j ]=i\delta_{ij}/2$ with $\hbar = 1/2$;
the subscripts $i$ and $j$ refer to the $i$-th and $j$-th modes.
In quantum optics, these observable represent the electric-field amplitudes in two orthogonal quadratures of each spatial mode.

CV cluster states are defined via the following stabilizer combinations~\cite{clusterPeter,Zhang2006} 
(so-called nullifiers~\cite{one-wayGu}):
\begin{align}
\left( \op_i - \sum_{j \in N_i} \ox_j \right),\ &\forall i \in G.
\label{eq:ideal cluster}
\end{align}
Here, nullifiers mean that the states become zero eigenstates of these quadrature combinations in the limit of infinite squeezing.
The modes of $i \in G$ correspond to the nodes of the graph $G$, while the modes of $j \in N_i$ are the nearest neighbors of the $i$-th mode.
The cluster state can be created by preparing $\ket{p=0}$ for each mode of $i \in G$ and performing QND interactions with modes of $j \in N_i$,
where the unitary operator for the QND interaction is $\hat{U}_{ij}=\exp (2i\ox_i \ox_j )$~\cite{Braginsky}.

With finite squeezing, the quadrature correlations are imperfect.
As sufficient conditions for cluster-type entanglement, we can use the following inequalities:
\begin{align}
\Bigl \langle \Delta (\op_i -\sum_{j \in N_i} \ox_j )^2 \Bigr \rangle &< \frac{1}{2}, &\forall i \in G,
\label{eq:cluster criteria}
\end{align}
because, for any $i \in G$ and $j \in N_i$, we can derive,
\begin{align}
\Bigl \langle \Delta (\op_i - \ox_j - \sum_{k \in N_{i|j}} \ox_k )^2 \Bigr \rangle + \Bigl \langle \Delta (\op_j - \ox_i - \sum_{l \in N_{j|i}} \ox_l )^2 \Bigr \rangle &< 1,
\end{align}
where $N_{i|j}$ denotes the set of all neighbor modes of the $i$-th mode except for the $j$-th mode.
This equation proves that the $i$-th mode cannot be in a state separable from the mode set $N_i \cup N_{j|i}$, according to 
the criteria of Ref.~\cite{vanLoock Furusawa}.

We can reverse the QND interaction and restore one of two {\em unknown} input states by using homodyne measurements and feedforward~\cite{Filip03.pra,erasing_paper}.
This restoration also does not depend on the specific interaction strength of the QND interaction.
In this case, the inverse QND interaction between the $i$-th and $j$-th modes $\hatd{U}_{ij}=\exp (-2i\ox_i \ox_j )$ can be decomposed into a homodyne measurement of the $x$-quadrature of the $j$-th mode with a subsequent phase-space displacement $\hat{Z}_i(x_j)=\exp (-2i\ox_i x_j)$ as feedforward to the $i$-th mode, where $x_j$ is the measurement result at the $j$-th mode.
Note that this measurement-based reversibility is specific to QND, controlled-NOT, and controlled $\pi$-phase-shift interactions. 

We consider removing the $j$-th mode from the cluster state through the erasing technique (Fig.~\ref{fig:removing}).
In this case, $x$-quadrature of the $j$-th mode is measured and feedforward to any mode of $i \in N_j$ is performed.
In the Heisenberg picture, the operators after the feedforward are $\op^\prime_i = \hatd{Z}_i(-x_j) \op_i \hat{Z}(-x_j)= \op_i - x_j$ for each $i \in N_j$.
As a resulting state, we obtain,
\begin{align}
\begin{cases}
\Bigl \langle \Delta (\op^\prime_i -\sum_{k \in N_{i|j}} \ox_k )^2 \Bigr \rangle &< \frac{1}{2} \text{, for } i \in N_j, \\
\Bigl \langle \Delta (\op_i -\sum_{k \in N_i} \ox_k )^2 \Bigr \rangle &< \frac{1}{2} \text{, for } i \in G^\prime -N_j,
\end{cases}
\end{align}
where $G^\prime = G|j$.
Thus, Ineq.~\eqref{eq:cluster criteria} is preserved even after the removal operations.
Therefore, the $j$-th mode is removed from the graph $G$ and cluster-type entanglement remains present among the resulting graph $G^\prime$.


The same technique can be also used for removing modes while still preserving the connections or entanglement of their neighbors, so-called ``wire shortening''~\cite{one-wayGu}.
Here, we consider the modes 1,2,3, and 4 in a cluster state constituting a wire from 1 to 4, $N_2 = \{1,3\}$ and $N_3 = \{2,4\}$ while the first mode and the fourth mode are not neighbors, as shown in Fig.~\ref{fig:wire_shortening_theory}.
The wire shortening requires that there are no other neighbor modes of the second and the third modes, but this requirement can be easily met by removing such modes prior to the wire shortening.
With local phase shifts of $\{\pi, -\pi/2, \pi/2, 0\}$ on each $\{1, 2, 3, 4\}$-th mode, the wire becomes a ring~\cite{4cluster}.
Thus, shortening of this wire corresponds to removing the second and third modes from the ring.
In the case of the ring, this operation is done by measuring $x$-quadratures of the second and third modes, and perform feedforward to the fourth and first modes, $\hat{Z}_4(-x_2)$ and $\hat{Z}_1(-x_3)$.
In the case of the wire, the operation is equivalent to measuring $\op_2$ and $\op_3$, and performing $\hat{Z}_4(-p_2)$ and $\hat{Z}_1(-p_3)$; as a resulting state, we obtain the cluster state with nullifiers $\left( \op_1 + \ox_4 - \sum_{i \in N_{1|4}} \ox_i \right)$ and $\left( \op_4 + \ox_1 - \sum_{j \in N_{4|1}} \ox_j \right)$.
Thus, the first and fourth mode are directly connected in the resulting state.
Note that the signs of $\ox_1$ and $\ox_4$ are opposite compared with Eq.~\eqref{eq:ideal cluster} because of the local phase shifts from the wire to the ring.
With these two transformations--removing unwanted modes and wire-shortening, we can generate many desired cluster states from a sufficiently large two-dimensional lattice~\cite{one-wayGu}, as shown in Fig.~\ref{fig:cluster}.

\begin{figure}[tb]
\centering
\subfigure[Removing an edge mode from a four-mode cluster state.]{
\includegraphics[clip, scale=0.3]{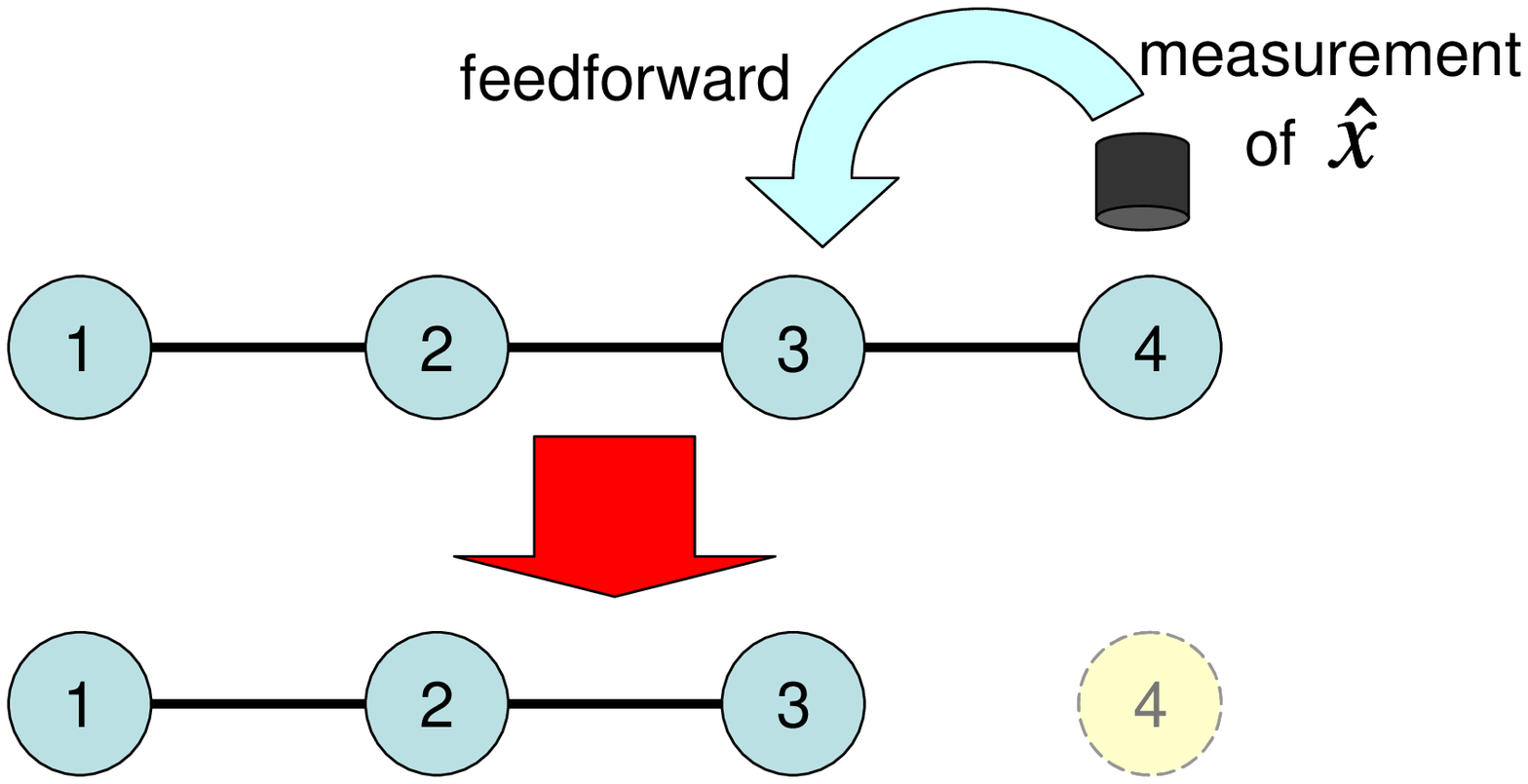}
\label{fig:removing4th}
}
\subfigure[Removing an inward mode from a four-mode cluster state.]{
\includegraphics[clip, scale=0.3]{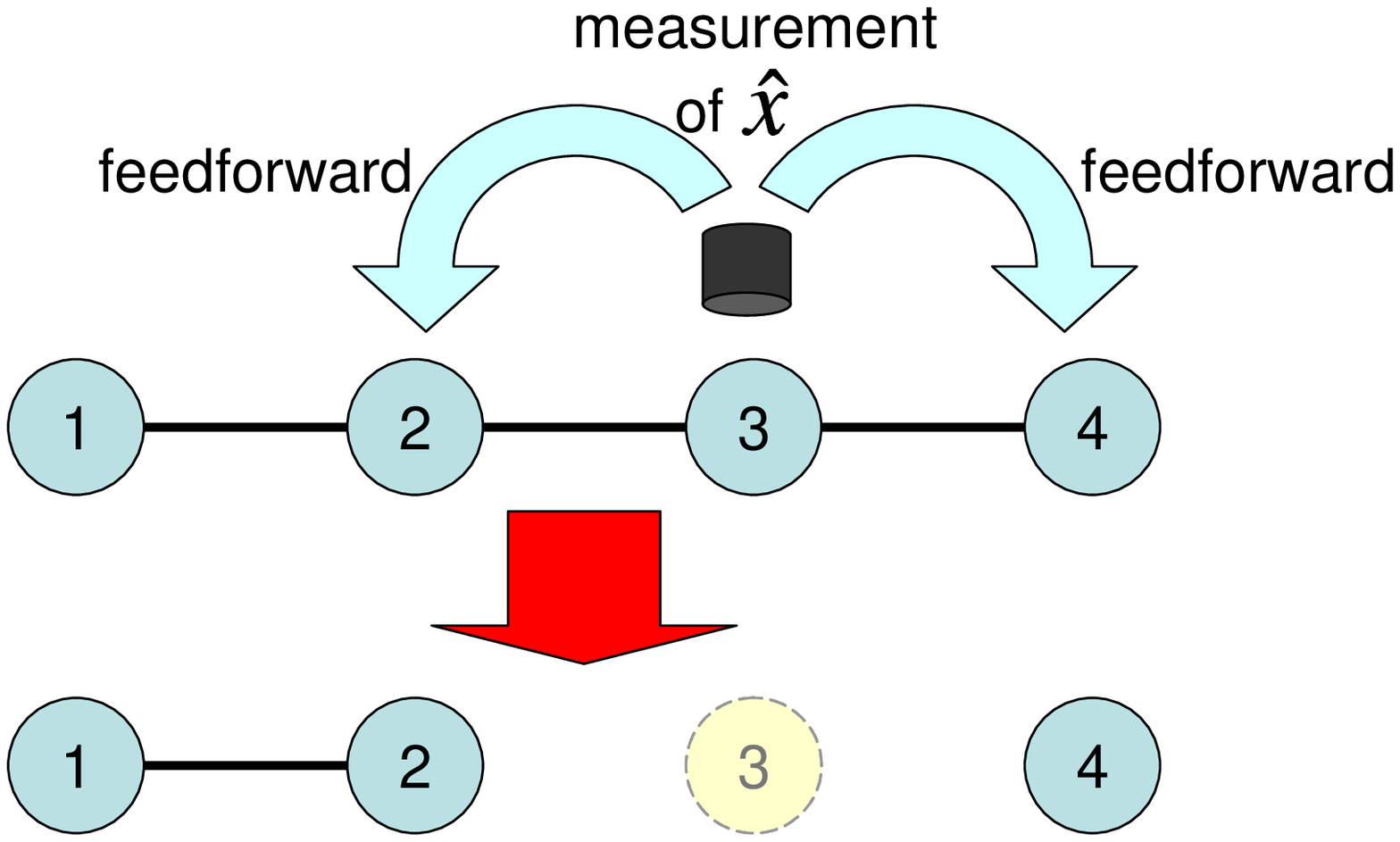}
\label{fig:removing3rd}
}
\subfigure[Wire-shortening from a four-mode cluster state.]{
\includegraphics[clip, scale=0.3]{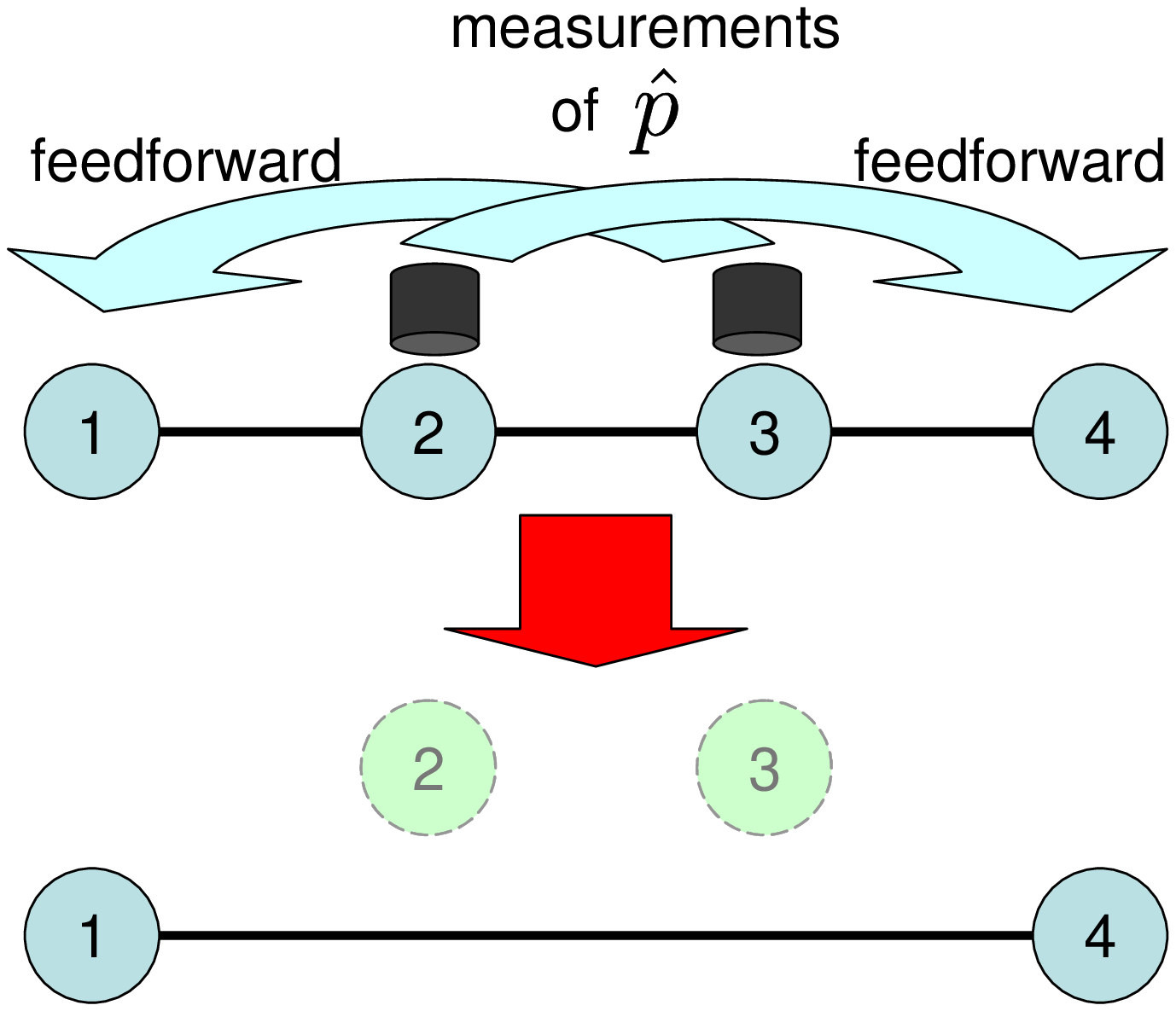}
\label{fig:wire_shortening2nd3rd}
}
\subfigure[Experimental setup for removing an edge mode from a four-mode cluster state.]{
\includegraphics[clip, scale=0.34]{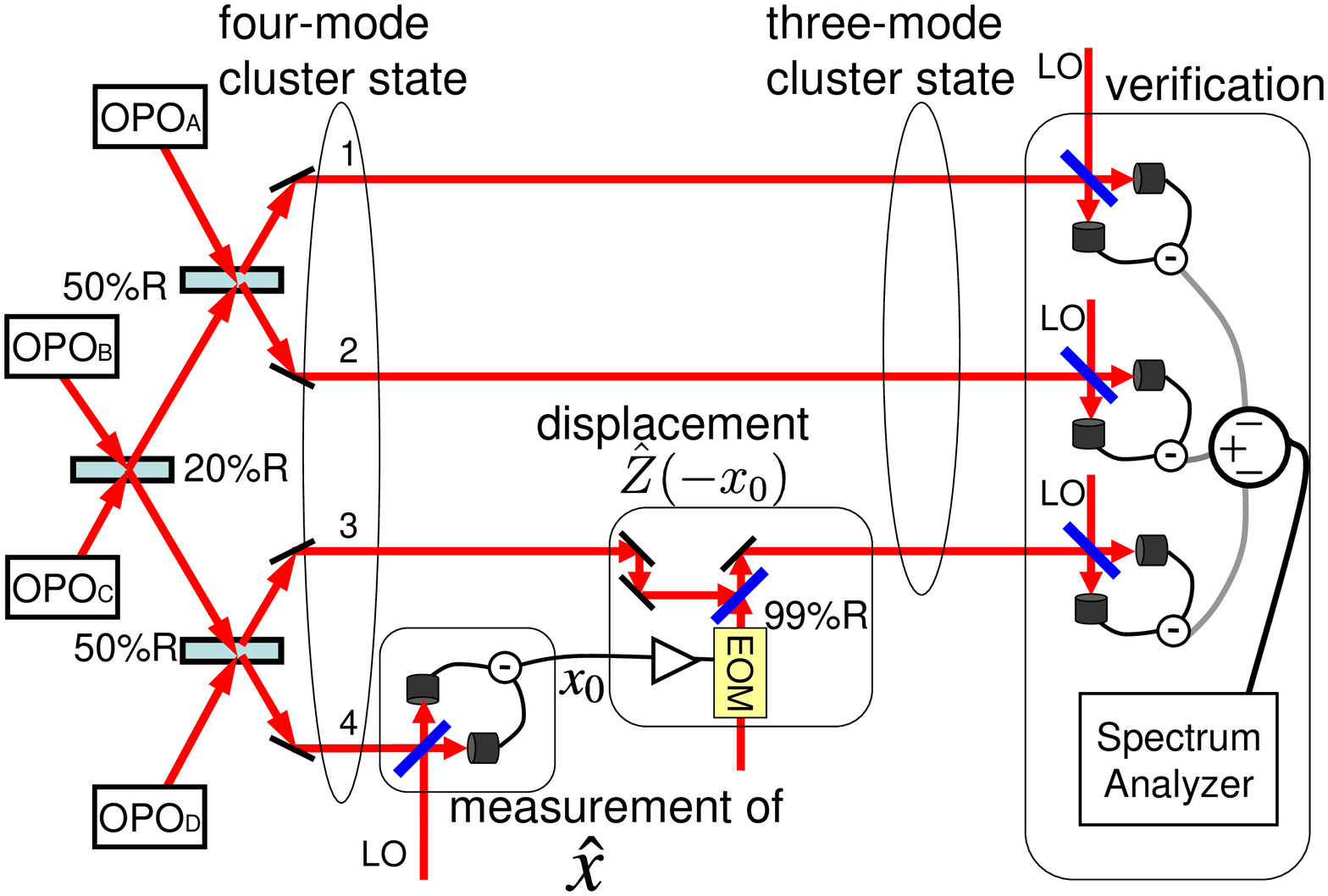}
\label{fig:cluster_setup}
}
\caption{Schematic of cluster state shaping and our experimental implementation. OPO: sub-threshold optical parametric oscillator generating a squeezed vacuum state, and LO: optical local oscillator for homodyne detection.}
\label{fig:setup}
\end{figure}

We experimentally demonstrate the above shaping operations on a four-mode linear cluster state.
Two of the following three experiments correspond to the removal of a mode, either at the edge of the cluster or within the cluster state (Fig.~\ref{fig:setup}[(a),(b)]).
The other experiment is wire-shortening (Fig.~\ref{fig:wire_shortening2nd3rd}).

\textit{Experimental Setup.}---Figure~\ref{fig:cluster_setup} is schematic of our experimental implementation of Fig.~\ref{fig:removing4th}.
This setup consists of generating a four-mode cluster state, shaping via measurement and feed-forward, and verification measurement.
As a light source, we utilize a continuous-wave Ti:sapphire laser with the wave length of 860~nm.
Four squeezed-vacuum states are generated from four sub-threshold optical parametric oscillators (OPOs).
Each OPO is a bow-tie shaped cavity of 500~mm in length with a 10-mm-long PPKTP crystal as a nonlinear medium~\cite{Suzuki06.apl},
which is pumped by the second harmonic (430~nm in wavelength) of Ti:sapphire output.
We generate a four-mode linear cluster state from four squeezed-vacuum states using three beam splitters with beam-splitting ratio of 20:80, 50:50, and 50:50 respectively~\cite{clusterPeter, 4cluster}.
In this experiment, $\pm1$ MHz sidebands are quantum modes while 98~kHz, 138~kHz, and 220~kHz modulation are used as phase references.
Squeezing level of resource squeeze vacuum states are about $-5$ dB.
The quantum correlation of the initial cluster states satisfies Ineq.~\eqref{eq:cluster criteria},
\begin{align}
\Bigl \langle \Delta (\op_i -\sum_{j \in N_j} \ox_b )^2 \Bigr \rangle &< 0.25 \pm 0.01 < \frac{1}{2},\ \ \text{for } i \in {1,2,3,4},
\label{eq:initial cluster}
\end{align}
thus, any mode of this cluster state is inseparable from the other modes.

In shaping operations, the modes being removed are measured via homodyne detection.
In order to perform feedforward, the electric signal of the detection outcome is amplified and drives an electro-optical modulator (EOM) traversed by an auxiliary beam with the power of 200~$\mu$W,
which is subsequently coupled with the neighbor mode of the measured mode by an asymmetric beam splitter (99:1).

Modes to be measured and to be suffered from feedforward depend on the shaping (Fig.~\ref{fig:setup}[(a)-(c)]).
For verification, remaining modes are measured via other homodyne detections.
The measurement-outcome electric-signals are combined and then sent to a spectrum analyzer in order to check the correlations between output quadratures.
The powers of the LOs are about 5~mW.
The detector's quantum efficiencies are greater than 99\%, and the interference visibilities to the LOs are on average 96\%.


\begin{figure}[tb]
\begin{tabular}{c}
\includegraphics[clip, scale=0.3]{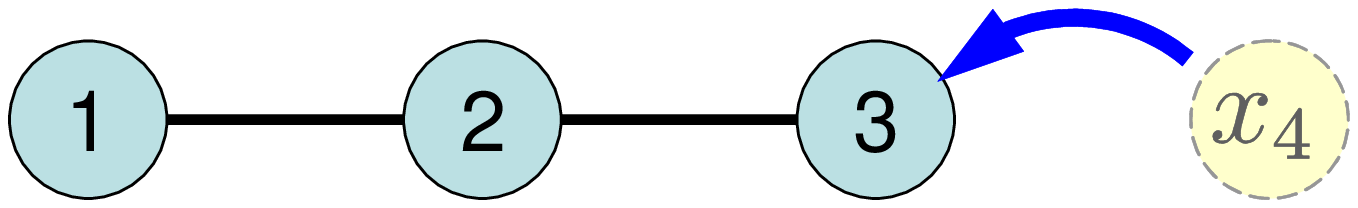}\\
\subfigure[$\expectation{(\op_1 -\ox_2)^2}$.]{
\includegraphics[clip, scale=0.4]{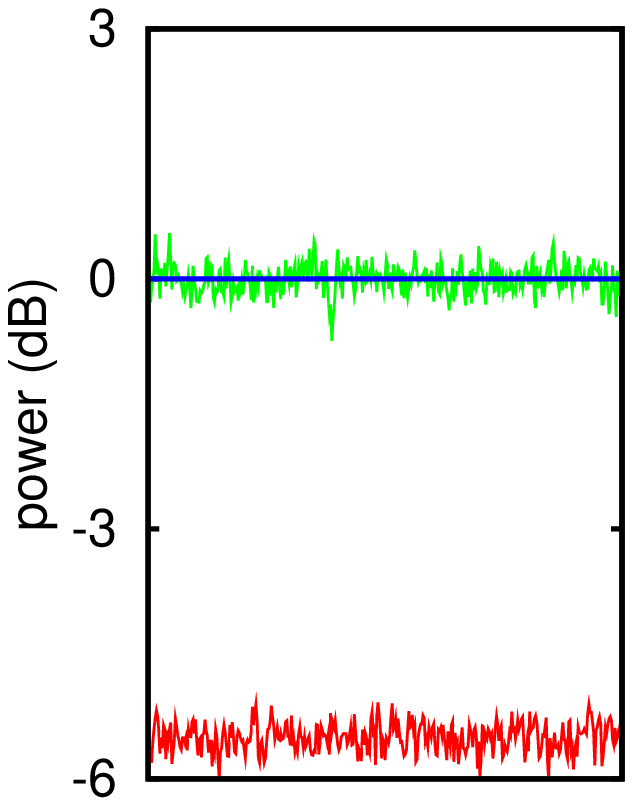}
}
\subfigure[$\expectation{(\op_2 -\ox_1 -\ox_3)^2}$.]{
\includegraphics[clip, scale=0.4]{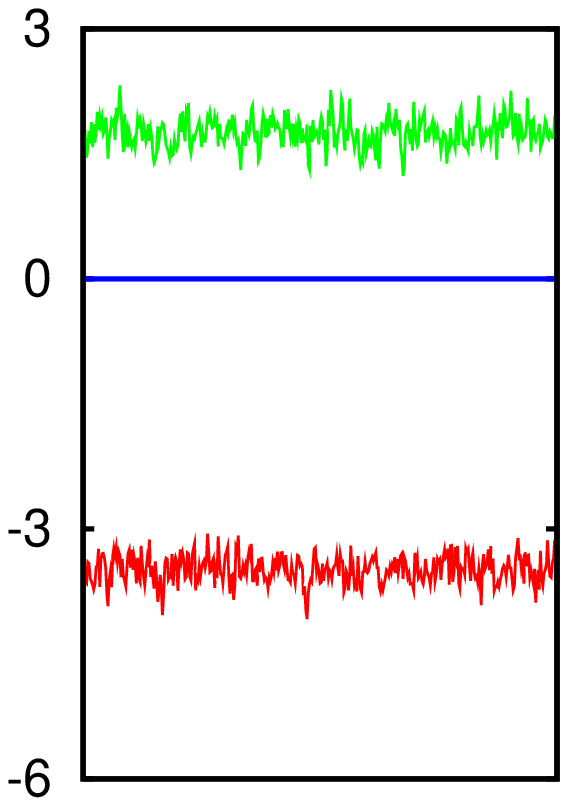}
}
\subfigure[$\expectation{(\op_3 -\ox_2)^2}$.]{
\includegraphics[clip, scale=0.4]{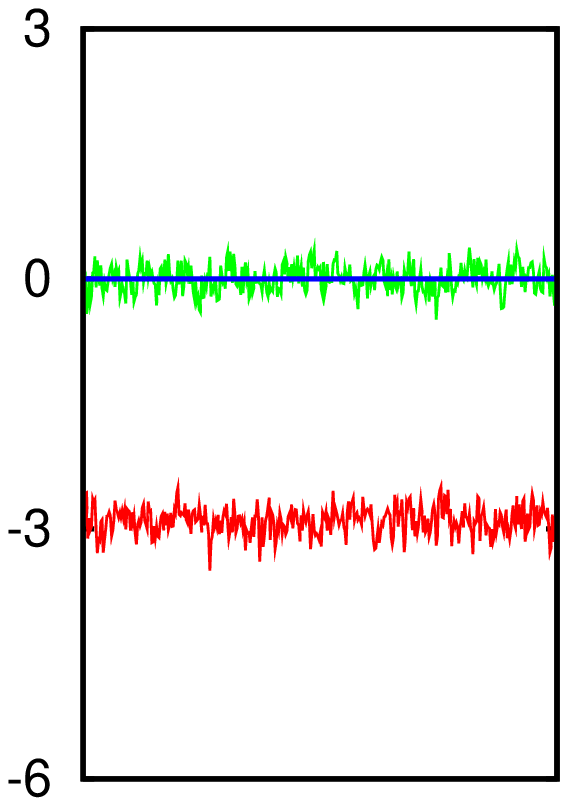}
}\\ \hline
\includegraphics[clip, scale=0.3]{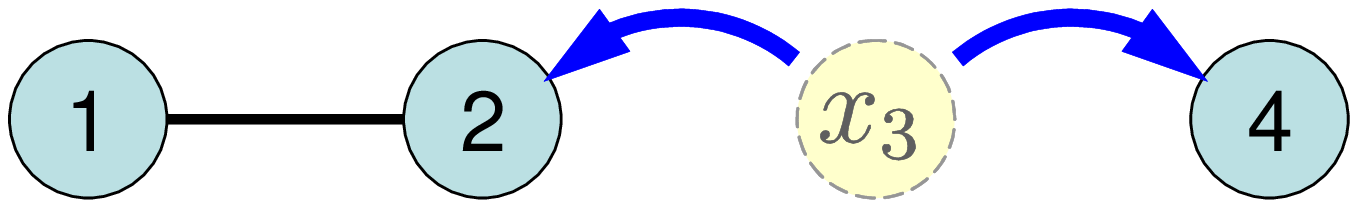}\\
\subfigure[$\expectation{(\op_1 -\ox_2)^2}$.]{
\includegraphics[clip, scale=0.4]{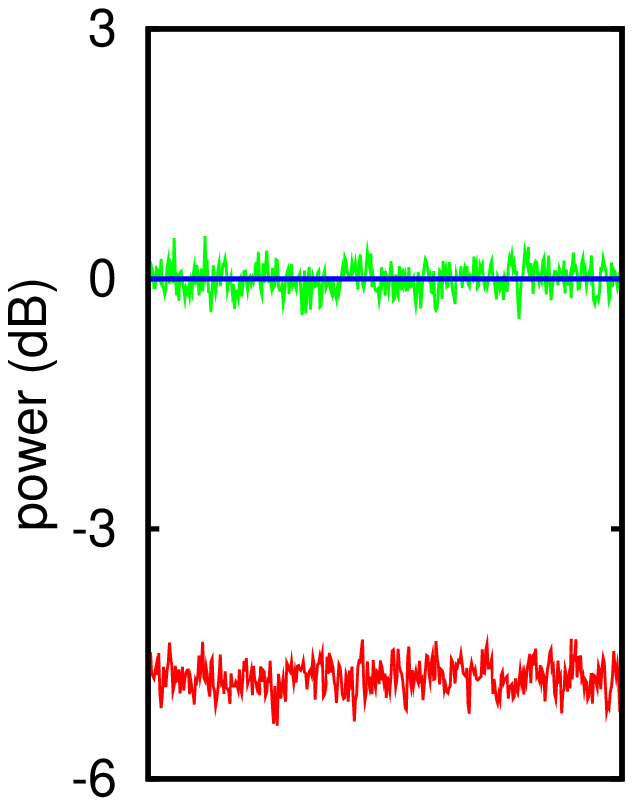}
}
\subfigure[$\expectation{(\op_2 -\ox_1)^2}$.]{
\includegraphics[clip, scale=0.4]{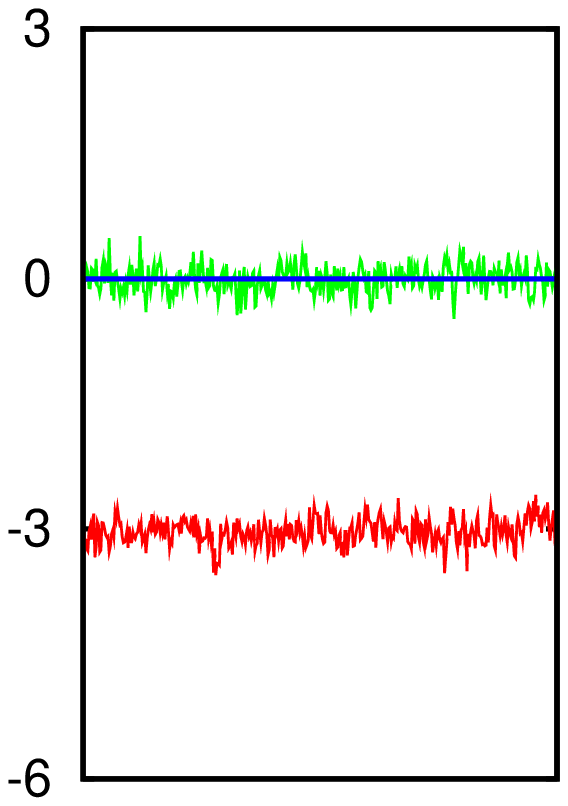}
}
\subfigure[resulting state of the fourth mode.]{
\includegraphics[clip, scale=0.4]{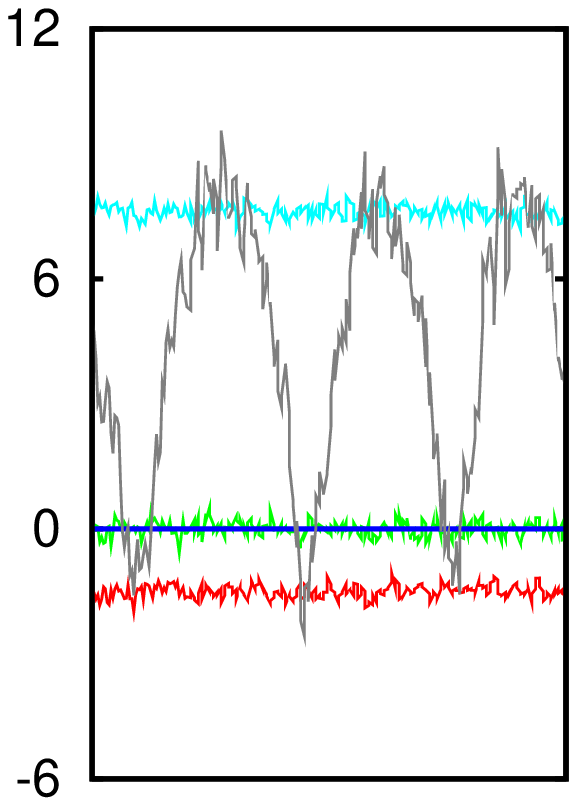}
}\\ \hline
\includegraphics[clip, scale=0.3]{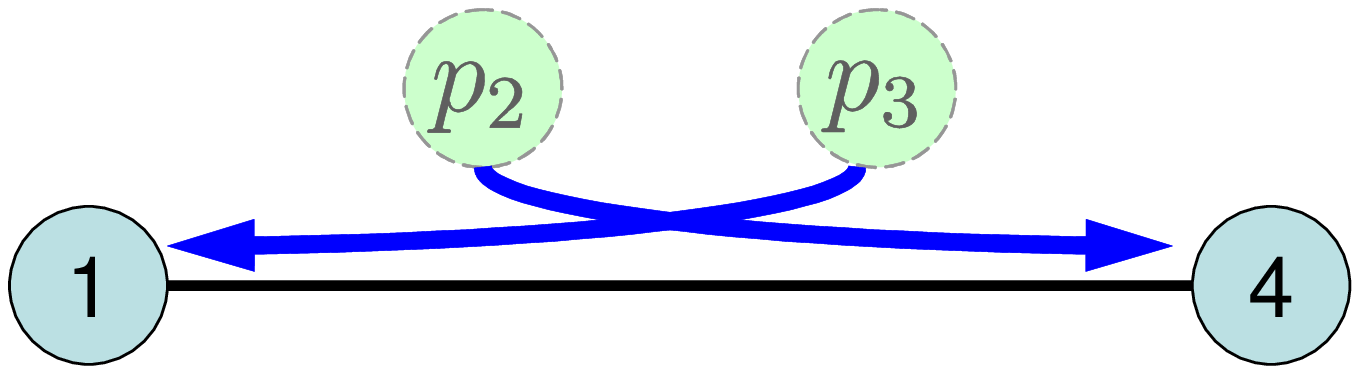}\\
\subfigure[$\expectation{(\op_1 +\ox_4)^2}$.]{
\includegraphics[clip, scale=0.4]{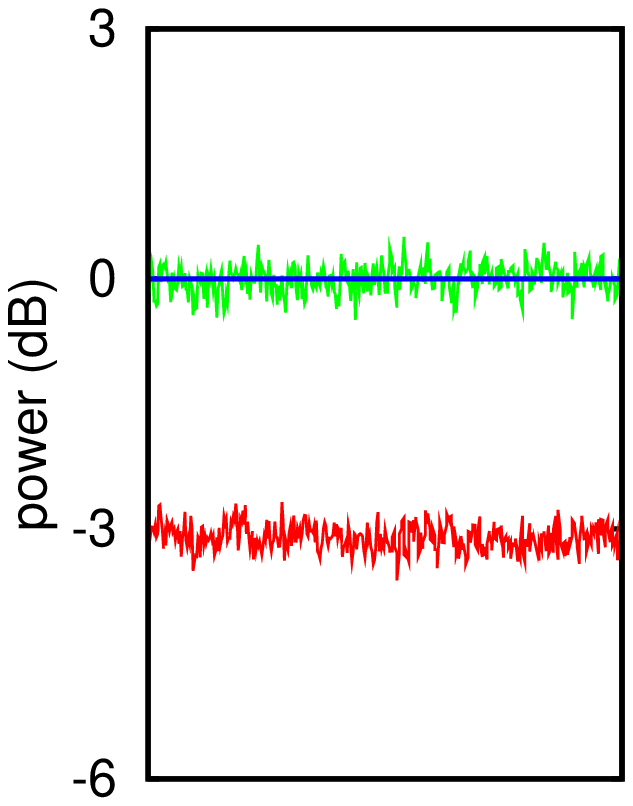}
}
\subfigure[$\expectation{(\op_4 +\ox_1)^2}$.]{
\includegraphics[clip, scale=0.4]{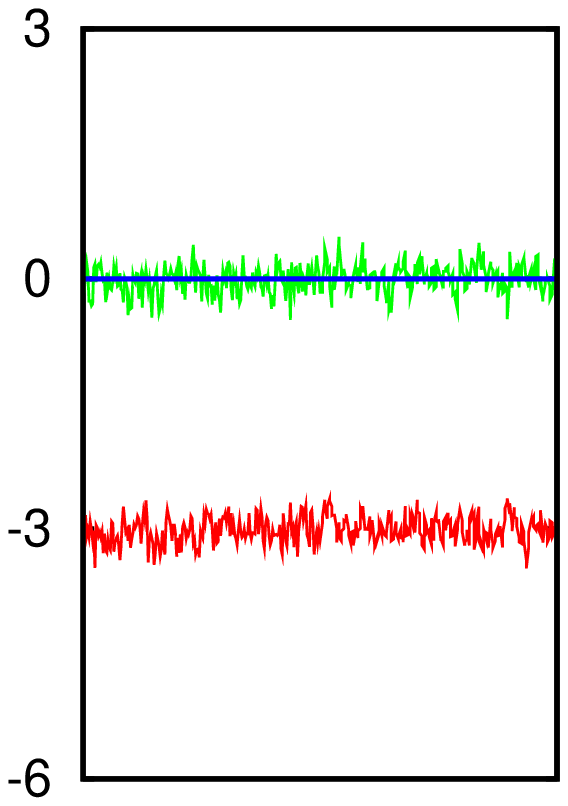}
}
\end{tabular}
\caption{Experimental results. [(a), (b), and (c)]: Quantum correlations of remains three modes in experiment represented in Fig.~\ref{fig:removing4th}.
Experimental results (red traces) are compared with vacua states (green traces), relative to two SNLs ($=1/2$, blue lines).
[(d) and (e)] shows remaining quantum correlations between first and second modes in experiment of Fig~\ref{fig:removing3rd}, while (f) shows the resulting state of the fourth mode. Variance of the squeezed quadrature (red trace), anti-squeezed quadrature (cyan), and observed variance with LO phase scanned (gray) are compared with vacuum state (green).
[(g) and (h)]: Reconstructed quantum correlations in wire-shortening experiment, represented in Fig.~\ref{fig:wire_shortening2nd3rd}.}
\label{fig:cluster_results}
\end{figure}

\textit{Experimental results.}---First, we demonstrate removal of the edge fourth mode from a four-mode linear cluster state, as shown in Fig.~\ref{fig:removing4th}.
Here, we measure $x$-quadrature of the fourth mode and perform feedforward to the third mode, $\hat{Z}_3(-x_4)$; and verify entanglement among the remaining modes.
Figure~\ref{fig:cluster_results}[(a)-(c)] shows experimental results of variances of nullifiers (red traces) suppressed below the cluster criteria Ineq.~\eqref{eq:cluster criteria} (blue lines),
\begin{align}
\begin{cases}
\expectation{\Delta (\op_1 -\ox_2 )^2} &= 0.14 \pm 0.01 < \frac{1}{2}, \\
\expectation{\Delta (\op_2 -\ox_1 -\ox_3 )^2} &= 0.22 \pm 0.01 < \frac{1}{2}, \\
\expectation{\Delta (\op_3 -\ox_2 )^2} &= 0.26 \pm 0.01 < \frac{1}{2}.
\end{cases}
\end{align}
Therefore, we successfully remove the fourth mode with preserving entanglement among the remaining modes.

Next, we also remove the inward third mode from the four-mode cluster state, as shown in Fig.~\ref{fig:removing3rd}.
This time, we measure $x$-quadrature of the third mode and perform feedforward to the second and fourth modes, $\hat{Z}_2(-x_3)$ and $\hat{Z}_4(-x_3)$.
Theoretically, the resulting states become a two-mode cluster state (the first and the second mode) and a squeezed state (the fourth mode).
As Figs.~\ref{fig:cluster_results}[(d) and (e)] show, we observed quantum correlations between the first and the second mode.
\begin{align}
\begin{cases}
\expectation{\Delta (\op_1 -\ox_2)^2} &= 0.17 \pm 0.01 < \frac{1}{2}, \\
\expectation{\Delta (\op_2 -\ox_1)^2} &= 0.25 \pm 0.01 < \frac{1}{2}.
\end{cases}
\end{align}
Since they satisfy Ineq.~\eqref{eq:cluster criteria}, these modes are entangled.
Meanwhile, the fourth mode recovers its squeezed property, as shown in Fig.~\ref{fig:cluster_results}(f).
Variance of squeezed quadrature is -1.5 $\pm$ 0.2 dB relative to the SNL.
Thus, a two-mode cluster state and a squeezed vacuum are obtained from a four-mode cluster state.



Finally, we perform wire shortening, as shown in Fig.~\ref{fig:wire_shortening2nd3rd}.
Again, we start with the four-mode cluster state, and then, measure $p$-quadratures of the second and third modes and perform feedforward to the fourth and first modes, $\hat{Z}_4(-p_2)$ and $\hat{Z}_1(-p_3)$.
Figure.~\ref{fig:cluster_results}[(g) and (h)] show quantum correlations of the resulting state.
The reconstructed quantum correlations are,
\begin{align}
\begin{cases}
\expectation{\Delta (\op_1 +\ox_4)^2} &= 0.25 \pm 0.01 < \frac{1}{2}, \\
\expectation{\Delta (\op_4 +\ox_1)^2} &= 0.24 \pm 0.01 < \frac{1}{2},
\end{cases}
\end{align}
satisfying Ineq.~\eqref{eq:cluster criteria}.
Therefore, two modes are removed while their quantum correlations are preserved between their neighbors.


Observable quantum correlations in Fig.~\ref{fig:cluster_results}~[(c), (e), (g) and (h)] are degraded by about 2 dB from the original resource squeezing levels of $-5$ dB.
The degradation correspond to a cost of top-to-bottom approach because generating a large cluster state requires extra QND interactions which are imperfect due to finite-squeezed resources~\cite{Yoshikawa08.prl}.
Nonetheless, technological progress toward increasing the experimental squeezing levels~\cite{sqrecord1,sqrecord2} will improve the cluster state shaping.

In conclusion, we have demonstrated experimentally how to remove an arbitrary node from a continuous-variable cluster state and how to shorten any quantum wires of such a state, where both transformations performed in unconditional fashions via quantum erasure.
In our experiment, all variations of this cluster-shaping have been demonstrated on a four-mode linear cluster state.
These two transformations can provide flexibility of fixed large-scale cluster states and transform them into appropriate forms for desired quantum operations.


This work was partly supported by SCF, GIA, G-COE, PFN and FIRST commissioned by the MEXT of Japan, the Research Foundation for Opt-Science and Technology, SCOPE program of the MIC of Japan, and ASCR-JSPS.
P.~v.~L. acknowledges support from the Emmy Noether programme of the DFG in Germany. R.~F. acknowledges projects: MSM
6198959213 and ME10156 of the Czech Ministry of Education, grant 202/08/0224 of GA \v CR and EU Grant
FP7 212008 COMPAS.

\end{document}